\documentclass[apj]{emulateapj}
\usepackage{amssymb}
\usepackage{bm}
\usepackage{amsmath}
\usepackage{color}
\usepackage{pst-grad}
\usepackage{psfrag}
\usepackage{graphicx}

\newcommand{\eqb}{\begin{eqnarray}}
\newcommand{\eqe}{\end{eqnarray}}
\newcommand{\diff}{{\rm d}}
\newcommand{\LW}{Li\'enard-Wiechart}
\newcommand{\bfm}[1]{\mbox{\boldmath$ #1 $}}
\newcommand{\im}{\rm i}
\newcommand{\intff}{\int_{-\infty}^{\infty}}

\newcommand{\curvature}{\kappa}
\newcommand{\torsion}{\bar{\tau}}

\begin{document}
\title{Computation of synthetic spectra from simulations of 
relativistic shocks} 
\author{
Brian Reville
and John G. Kirk}
\affil{Max-Planck-Institut f\"ur Kernphysik, Postfach 10~39~80,
69029 Heidelberg, Germany}
\email{brian.reville@mpi-hd.mpg.de, john.kirk@mpi-hd.mpg.de}

\begin{abstract}

  Particle-in-cell (PIC) simulations of relativistic shocks are in
  principle capable of predicting the spectra of photons that are
  radiated incoherently by the accelerated particles. The most direct
method evaluates the spectrum using the fields given by the \LW~
  potentials. However, for relativistic particles this procedure is
  computationally expensive.  Here we present an alternative method,
  that uses the concept of the photon formation length. 
The algorithm is suitable for
  evaluating spectra both from particles moving in a specific realization
  of a turbulent electromagnetic field, or from trajectories given 
 as a finite, discrete time series by a PIC simulation. The main
  advantage of the method is that it identifies the intrinsic spectral 
features, and filters out those that are
  artifacts of the limited time resolution and finite duration of 
input trajectories.

\end{abstract}


\maketitle

\section{Introduction}

Particle acceleration at relativistic shocks is thought to be
responsible for the high-energy nonthermal photons observed from a
variety of astrophysical objects, such as gamma-ray bursts, pulsars
and blazars. To test this hypothesis, reliable predictions of the
photon spectra are needed. Analytic models have been used to provide
estimates of the expected asymptotic power-law index at high energy
and the maximum attainable photon energy
\citep{derishev07,kirkreville10}, but they cannot currently take account
of potentially important effects, such as the role of self-generated
turbulence in the vicinity of the shock.  Particle in cell (PIC)
simulations, on the other hand, have the potential to capture these effects,
and have recently begun to provide evidence that Fermi 
acceleration is a natural consequence
of relativistic shock formation
\citep{spitkovsky05,spitkovsky08a,spitkovsky08b,sironispitkovsky09a,martinsetal09}.
In principle, these simulations are capable of reproducing the
essential physics; they are {\em ab initio} in the sense that all
processes are reproduced 
by evolving the electromagnetic fields and the particle distribution
according to the
classical equations of motion and Maxwell's equations.

However, to compare the simulations with observations, it is essential
to understand the predicted radiative signatures.
Using results from PIC simulations, several groups have
computed the emission at relativistic shocks
\citep{hededalphd,jmartinsetal09,sironispitkovsky09b,medvedevetal10},
but the results show substantial differences. This is not necessarily
due to the way in which the spectra were evaluated, since the 
electromagnetic fields and energetic
particle distribution vary strongly from simulation to simulation. On the 
other hand, it does not rule out such a dependence. In each case, the method 
employed to compute the emission is the same: the electric field 
produced at a virtual detector by a single particle trajectory
is evaluated using the 
\LW~potentials, the result is Fourier transformed and then
averaged 
over a large family of trajectories. 
This procedure is expensive in terms of computing resources, 
 especially if one wants to compute the 
high-energy emission of relativistic particles, because 
of (i) the extremely high time resolution required to describe 
high-energy photons, (ii) the large number of virtual detectors 
required to 
resolve the narrow radiation beam of a relativistic particle 
(which scales as $\gamma^2$) 
and (iii)
the long time series needed to account for low-frequency emission.

In this paper, we present an alternative approach. 
For a given observing frequency, we identify at each point on 
a particle trajectory
the length that contributes coherently to the emission. 
In a quantum picture, this is known as the photon formation length
\citep{akhiezershulga87}. Using this as a guide, we then perform the integrations 
using a new algorithm that is optimized for highly relativistic particles. 

The essential information on the particle trajectory can be supplied to the 
algorithm in two different ways:
If the electromagnetic fields are prescribed
as a function of space and time,
then the trajectory can be integrated using standard adjustable-step methods.
As examples, we present in section~\ref{turbulentfields} 
computations of the 
emission spectrum from isotropic particle distributions immersed in 
stationary, turbulent magnetic fields. In this case, the algorithm
is employed in the inner loop of a multi-dimensional integration, which is 
performed using a Monte-Carlo method.
On the other hand, if the fields
are not known at all points in space and time, interpolation is required. 
This is the case, for example, in PIC simulations, where 
the fields, particle positions and velocities are known only 
at discrete times and locations. We discuss this situation and 
suggest a procedure for implementing the algorithm
in section \ref{PIC_sect}.

\section{Equations for the emissivity}
\label{sect_emissivity}

In the classical theory of electrodynamics,
the spectral and angular distribution of radiation produced by a single 
particle in vacuum in the direction $\bm{n}$ is given by the well-known 
formula \citep[e.g.][]{landaulifshitz}
\eqb
\label{LWrad}
\frac{\diff E}{\diff \omega \diff\Omega}=\frac{q^2}{4\pi^2c}
\left|\int_{-\infty}^{+\infty} \frac{\bm{n}\times \left[({\bf
          n}-\bm{\beta})\times \dot{\bm{\beta}}\right]}{(1-
	  \bm{n}\cdot\bm{\beta})^2} {\rm e}^{ \im(\omega t-{\bf k}\cdot
      {\bf \bm{x}}(t))} \diff t\right|^2 
\label{startingpoint}
\eqe 
where ${\bf k}=\omega\bm{n}/c$. 
Equation (\ref{LWrad}) is usually used as the starting point for the numerical 
computation of radiation signatures from PIC codes 
--- a detailed description of
the method can be found in
\citet{hededalphd}. 
However, there are three disadvantages of this form of the emissivity
\begin{enumerate}
\item
the term $\textrm{e}^{i\omega t}$ is rapidly oscillating
\item
the term $1-\bm{n}\cdot\bm{\beta}(t)$ in the denominator produces
a very sharply peaked function when used for the trajectory of a  
relativistic particle
\item
the range of integration extends over the entire section of the 
trajectory on which the acceleration is nonzero, making it difficult to relate 
the expression to a local emissivity and, hence, to compute
time-dependent emission.
\end{enumerate}

Straightforward transformations lead to a number of alternative forms for
Equation (\ref{LWrad}), for example, 
\eqb 
\frac{\diff E}{\diff \omega \diff\Omega}= \intff
P(\bm{n},\omega, t) \diff t
\label{LWenergy}
\eqe
where
\eqb
 P(\bm{n},\omega, t)&=&-\frac{q^2\omega^2}{4\pi^2c}
\intff\diff \tau 
\left[{1}-{\bfm{\beta}}(t+\tau)\cdot{\bfm{\beta}}(t)\right]\nonumber\\
&&\cos\left(\omega\left[\tau-\bm{n}\cdot(\bm{x}(t+\tau)-\bm{x}(t))/c\right]\right)
\enspace.
\label{pnot}
\eqe 
This expression is exact, and has the advantage
that, provided variations on the timescale $\omega^{-1}$ are small,
the quantity $P(\bm{n}, \omega,t)$, when suitably averaged, 
can be interpreted as the instantaneous spectral power
radiated per unit solid angle about the direction $\bm{n}$
\citep{schwinger49}. 

In practice, it is necessary to truncate the integrals 
in (\ref{LWenergy}) and (\ref{pnot}) to finite intervals. 
From the form of the integrand in (\ref{pnot}), it is clear that 
the endpoints
should be chosen such that at least the first few periods of the
cosine function are included. This leads to the 
concept of the photon formation time or coherence time, which 
applies to both the quantum and classical formulations of the problem
\citep[for a review, see][]{akhiezershulga87}.
For a given Fourier mode, with wavelength $\lambda=2\pi c/\omega$, 
a particle trajectory contributes coherently to the instantaneous power 
radiated at time $t$
until it has lagged at least one wavelength behind the 
wavefront emitted at time $t$. 
Thus, the coherence
or formation time $\tau_{\rm coh}$ is determined implicitly by the 
equation
 \eqb
\omega\left(\tau_{\rm coh}-\left|\bm{x}(t+\tau_{\rm coh})-
      \bm{x}(t)\right|/c\right)= 2\pi\enspace.
\label{cohtime}
\eqe 
To compute the radiated power, one needs to know
the trajectory accurately over
several coherence times.
For relativistic particles, a wavefront
can take a considerable amount of time to separate one wavelength from
the particle, particularly at low frequencies, when the wavelength is long.

In the context of Fermi acceleration at
relativistic shocks, an angular dependent calculation
of the emission from an individual particle is unnecessary, 
provided one is interested 
only in the high-energy emission from accelerated particles.
This is because the characteristic radiation
beaming angle of $1/\gamma$ is much smaller for these particles
than the scales on which
anisotropy in the particle distribution can be expected, which is roughly the
reciprocal of the Lorentz factor of the fluid motion into the shock 
\citep{achterbergetal01}.
Hence, when
summed over all plasma particles, these sharp emission peaks are smoothed out.  
In this case, it is advantageous
to work with an angle-integrated expression for the 
individual particle spectrum.
Integrating equation (\ref{pnot}) over
solid angle, gives:
\eqb
\frac{\diff E}{\diff \omega}=
\intff P(\omega, t) \diff t
\label{schwingerenergy}
\eqe
where
\eqb
P(\omega,t)&=&\frac{e^2\omega}{2\pi c}\int_{-\infty}^{\infty}\diff\tau\,
\left[1-\bm{\beta}(t)\cdot\bm{\beta}(t+\tau)\right]F(\omega,t,\tau)
\label{pot}
\eqe
with
\eqb
F(\omega,t,\tau)&=&
\frac{\sin\left[\omega\left(\tau-\Delta\right)\right]
-\sin\left[\omega\left(\tau+\Delta\right)\right]
}{\Delta}
\label{rapidlyosc}
\\
\Delta(t,\tau)&=&\left|\bm{x}(t+\tau)-\bm{x}(t)\right|/c
\label{defdelta}
\eqe 
Equation (\ref{pot}) is also exact, and $P(\omega,t)$ can be
interpreted as the power radiated at time $t$ in unit angular
frequency range, again subject to the condition that it varies slowly
on the timescale $\omega^{-1}$ \citep{schwinger49}.  This condition is
not always fulfilled for the trajectories we consider. In particular,
it is violated 
when the acceleration felt by the particle fluctuates rapidly whilst
the velocity remains within the beaming angle of the radiation (\lq\lq
jitter\rq\rq\ radiation).  Nevertheless, equation
(\ref{schwingerenergy}) for the total radiated energy remains valid,
although $P(\omega,t)$, which we call the \lq\lq instantaneous
power\rq\rq, is not necessarily positive definite.

\subsection{Computation of the instantaneous power}
\label{sect_algorithm}
 
For relativistic particles, and for small $\tau$, such that 
the particle displacement
$\Delta$ defined in (\ref{defdelta}) is approximately $\beta|\tau|$,
the function $F(\omega,t,\tau)$ in (\ref{rapidlyosc}) 
contains two kinds of term: those that oscillate rapidly in 
$\tau$ with frequency $\sim\omega$, and those that oscillate slowly,
with frequency $\omega/\gamma^2$. Physically, the latter arise because the
particle chases the wavefront, remaining close to it for a relatively 
long time. It is convenient to separate these terms:
\eqb
P(\omega,t)&=&P_1(\omega,t)+P_2(\omega,t)
\label{ptotal}
\\
P_1(\omega,t)&=&
\frac{e^2\omega}{2\pi c}\int_{-\infty}^{\infty}\diff\tau\,
\left[1-\bm{\beta}(t)\cdot\bm{\beta}(t+\tau)\right]
\nonumber\\&\,&
\frac{
\sin\left[\omega\tau\left(1-\frac{\Delta}{|\tau|}\right)\right]
}
{\tau \Delta/|\tau|}
\label{p1}
\\
P_2(\omega,t)&=&
-\frac{e^2\omega}{2\pi c}\int_{-\infty}^{\infty}\diff\tau\,
\left[1-\bm{\beta}(t)\cdot\bm{\beta}(t+\tau)\right]
\nonumber\\&\,&
\frac{
\sin\left[\omega\tau\left(1+\frac{\Delta}{|\tau|}\right)\right]}
{\tau\Delta/|\tau|}
\label{p2}
\eqe
In the case of $P_2$, the 
time $\tau$ over which the trajectory is sampled is very short,
$\sim1/\omega$, whereas in the case of $P_1$ it is 
much longer, $\sim\gamma^2/\omega$.
To develop an approximation scheme for these terms we introduce 
quantities 
that describe the deviation 
of the trajectory from ballistic motion. These are
the deviation in position
\eqb
\delta\bm{x}(t,\tau)&=&\bm{x}(t+\tau)-\bm{x}(t)-c\tau\bm{\beta}(t)\,,
\label{devposition}
\\
\noalign{\hbox{the deviation in velocity:}}
\delta\bm{\beta}(t,\tau)&=&\bm{\beta}(t+\tau)-\bm{\beta}(t)
\label{devvelocity}
\\
\noalign{\hbox{and the deviation of the displacement}}
\delta\Delta(t,\tau)&=&\Delta(t,\tau)- |\tau|\beta(t)\,.
\label{devdisplacement}
\eqe

Clearly, to zeroth order in these deviations, the instantaneous 
power must vanish, since a particle undergoing uniform motion does
not radiate. Furthermore, because of the relatively long sampling time,
the dominant higher order contributions in the deviations come from $P_1$.
For frequencies large compared to the instantaneous angular frequency
(the local gyrofrequency), the higher order contributions in 
the $P_2$ term can be neglected to give
\eqb
P(\omega,t)&\approx&
\frac{e^2\omega}{2\pi c}
\int_{-\infty}^\infty\diff \tau\left\lbrace
\,\left[1-\beta^2(t)-\bm{\beta}(t)\cdot
\delta\bm{\beta}(t,\tau)\right]\right.\nonumber\\&\,&\left.
\frac{
\sin g(\omega,t,\tau)}{\tau\left[\beta(t)+\delta\Delta(t,\tau)/|\tau|\right]}
\right.\nonumber\\
&\,&\left.
-
\left[1-\beta^2(t)\right]
\frac{
\sin \left[\omega\tau\left(1+\beta(t)\right)\right]}{\tau\beta(t)}
\right\rbrace
\label{firstorderint}
\eqe
where we have introduced the phase-lag
$g(\omega,t,\tau)$:
\eqb
g(\omega,t,\tau)&=&\omega\tau\left[1-\beta(t)-
\frac{\delta\Delta(t,\tau)}{|\tau|}\right]
\label{phaselagdef}
\eqe

At this point it would be possible to proceed by evaluating 
analytically the integral involving the second term in 
(\ref{firstorderint}):
\eqb
\frac{1}{\gamma^2(t)\beta(t)}\int_{-\infty}^\infty\diff\tau\,
\frac{\sin\left[\omega\tau\left(1+\beta(t)\right)\right]}{\tau}
&=&\frac{\pi}{\gamma^2(t)\beta(t)}\enspace. 
\label{dirichlet}
\eqe
Indeed, \citet{schwinger49} followed this path in deriving 
an analytic expression for the synchrotron emissivity. 
However, because the integrands are oscillatory,
it is instead preferable to group them together. Transforming
the integration
variable from $\tau$ to the phase-lag $g$ defined in (\ref{phaselagdef}),
in the case of the first term in (\ref{firstorderint}), and as
$g=\omega\tau(1+\beta(t))$ in the case of the second term, leads to
\eqb
P(\omega,t)&=&
\frac{e^2\omega}{2\pi c}
\int_{-\infty}^\infty
\diff g\,\frac{\sin g}{g}
\nonumber\\
&&
\left\lbrace
\left[\frac{1}{\gamma^2(t)\beta(t)}\right]
\left[\frac{\beta(t)g}{\tau\dot{g}\left[\beta(t)+\frac{\delta\Delta(t,\tau)}{|\tau|}\right]}-1\right]
\right.
\nonumber\\
&-&\left.\frac{
g\bm{\beta}(t)\cdot\delta\bm{\beta}(t,\tau)}
{\dot{g}\tau\left[\beta(t)+\frac{\delta\Delta(t,\tau)}{|\tau|}\right]}
\right\rbrace
\label{instpower}
\eqe
where
\eqb
\dot{g}&=&\frac{\partial g}{\partial\tau}
\nonumber\\
&=&\omega-
\frac{\omega\bm{\beta}(t+\tau)\cdot\left[\bm{x}(t+\tau)-\bm{x}(t)\right]}
{c\tau\Delta(t,\tau)/|\tau|}
\\
&=&\omega-
\frac{\omega\left\lbrace
\left[\bm{\beta}(t)+\delta\bm{\beta}(t,\tau)\right]\cdot
\left[c\tau\bm{\beta}(t)+\delta\bm{x}(t,\tau)\right]\right\rbrace}
{c\tau\left[\beta(t)+\frac{\delta\Delta(t,\tau)}{|\tau|}\right]}
\label{gdotdefinition}
\eqe
As required, $P$ vanishes to zeroth order in the deviations
from a ballistic orbit, (\ref{devposition}), (\ref{devvelocity})
and (\ref{devdisplacement}). The grouping of the terms in
Equation (\ref{instpower}) in this manner is 
especially important at high frequencies, where
the higher order terms in $P_1$ and $P_2$
are small. In this limit, the two terms can be expressed as
\eqb
\lim_{\omega\rightarrow\infty} P_{1,2}=
\pm\frac{e^2\omega}{2 c}\frac{1}{\gamma(t)^2\beta(t)}
\eqe
and cancel exactly when summed. In a numerical evaluation, 
a small error remains, which grows linearly with $\omega$. Grouping the
terms together prevents the growth of this error.

Under the assumptions that the electromagnetic fields vary slowly on the 
timescale of a photon formation length, and that linear acceleration 
emission \citep[e.g.][]{schwinger49} is unimportant, we demonstrate 
in appendix~\ref{appendix_synchrotron} that (\ref{instpower}) reduces to 
a local emissivity. 
This is an obvious generalization of standard  
synchrotron emission, which takes 
account of acceleration in both
magnetic and electric fields by formulating it 
in terms of the local curvature of the trajectory:
\eqb
P(\omega,t)&=&
\frac{\sqrt{3}e^2\gamma\kappa}{2\pi}
\frac{\omega}{\omega_{\rm c}}\int_{\omega/\omega_{\rm c}}^\infty
\diff x\,K_{5/3}(x)
\label{instsyncheq}
\eqe
where 
\eqb
\omega_{\rm c}&=&3\gamma^3c\kappa/2
\eqe
and the curvature $\kappa$ is defined locally 
in terms of the particle velocity
and acceleration $\bm{\beta}$ and $\dot{\bm{\beta}}$:
\eqb
\kappa&=&
\frac{\left|\bm{\beta}\times\bm{\dot{\beta}}\right|}{c\beta^3}
\label{curvature_def}
\eqe
A perturbative approach that includes linear acceleration emission
as a first order correction to (\ref{instsyncheq})
has been presented by
\citet{melrose78}.

To perform the integration in (\ref{instpower}) numerically, we first 
split it at the points where $\sin g=0$, i.e., $g=n\pi$,
$(n=0,\pm1,\pm2\dots)$, and write it as an
infinite sum
\eqb
P(\omega,t)&=&
\frac{e^2\omega}{2\pi c}\sum_{n=-\infty}^{\infty}
\int_{n\pi}^{(n+1)\pi}
\diff g\,Q(g,t){\sin g}
\label{instpower2}
\\
\noalign{\hbox{where}} Q(g,t)&=&\frac{1}{g}\left\lbrace
  \left[\frac{1}{\gamma^2(t)\beta(t)}\right]
  \left[\frac{\beta(t)g}{\tau\dot{g}\left[\beta(t)+\frac{\delta\Delta(t,\tau)}{|\tau|}\right]}-1\right]
\right.
\nonumber\\
&&\,\,\,\,\,\,\,\,\,-\left.\frac{
    g\bm{\beta}(t)\cdot\delta\bm{\beta}(t,\tau)}
  {\dot{g}\tau\left[\beta(t)+\frac{\delta\Delta(t,\tau)}{|\tau|}\right]}
\right\rbrace 
\eqe 
and $\tau$ and $\dot{g}$ are considered to be
functions of $g$ and $t$, defined implicitly in (\ref{phaselagdef})
and (\ref{gdotdefinition}). According to its definition
(\ref{cohtime}), integration from $g=0$ to $g=2\pi$ corresponds
precisely to integration over one photon formation time. We therefore
anticipate on physical grounds that taking the first few terms 
should give a good approximation. However, the function 
$Q(g,t)$ grows linearly with $g$ for small $g$, before decreasing 
monotonically above some critical value $g^*$. In this case,
the sum in (\ref{instpower2}) 
does not begin to converge until $n>n^*=g^*/\pi$. 
For constant curvature,
it is straightforward to show that $g^*\approx3\omega/\omega_{\rm c}$,
so that $n^*$ becomes large only if one tries to compute the emissivity 
well above the cut-off frequency.
In general, we have found that a substantial improvement 
can be 
achieved by employing the
Euler--van Wijngaarden transform 
\cite[e.g.][section 5.1]{numericalrecipes}
to accelerate the convergence, whilst retaining a minimum number of 
about 20 terms in order to preserve accuracy at high frequencies, where the 
power radiated is low.

Evaluation of the instantaneous power based on (\ref{instpower2}) requires
knowledge of the functions $\beta(\tau)$ and $\delta\Delta(\tau)$. 
In the next section we apply this approach to finding the 
angular integrated emission of an isotropic, mono-energetic particle
distribution in prescribed, stationary, turbulent fields, in which 
these functions can be found using an adjustable-step integration of 
the trajectory. In 
section~\ref{PIC_sect}, on the other hand, we discuss the application 
to a trajectory that is known only as a discrete time series, for example, 
a trajectory from a PIC simulation.

\section{Prescribed fields}
\label{turbulentfields}

\subsection{Isotropic particle distribution}
\label{klim_sect}

Equation (\ref{schwingerenergy}) 
describes the energy emitted by a single particle.  If we now
consider the possibility of $N$ particles emitting incoherently whilst
following trajectories in a prescribed field in a volume $V$, 
and allow them to do so
for a time $T$, then the average power $L$ emitted by these particles
is obtained by summing over the individual contributions:
\eqb
\frac{\diff L}{\diff \omega}&=&\lim_{T\rightarrow\infty}
\sum_{i=1}^N\,\frac{1}{T}\int_{-T/2}^{T/2}\diff t P_i(t)
\eqe
where $P_i(t)$ is the instantaneous power of the $i$'th particle. 
Replacing
the sum by an integral over the exact (Klimontovich)
phase space distribution 
$f_{\rm K}(\bm{x},\bm{p},t)=\sum_{i=1}^N
\delta\left[\bm{x}-\bm{x}_i(t)\right]
\delta\left[\bm{p}-\bm{p}_i(t)\right]$
where $\bm{x}_i(t),\bm{p}_i(t)$
are the phase-space coordinates of the $i$'th particle at time $t$,
leads to 
\eqb
\frac{\diff L}{\diff\omega}&=&
\int\diff^3\bm{x}\,\diff^3\bm{p}\,
\frac{1}{T}\int_{-T/2}^{T/2}\diff t\,
f_{\rm K}(\bm{x},\bm{p},t)P(\bm{x},\bm{p},t)
\eqe
where $P(\bm{x}_i(t),\bm{p}_i(t),t)=P_i(t)$. 

In general, both the electromagnetic fields that determine 
the particle trajectories and the phase space distribution fluctuate 
in time. However, $L$ is 
a time-averaged quantity. If we are interested in the emission
from a system containing prescribed, static fields, then 
$P(\bm{x},\bm{p},t)$ is not an explicit function of time, so that 
\eqb
\frac{\diff L}{\diff\omega}&=&
\int\diff^3\bm{x}\,\diff^3\bm{p}\,
P(\bm{x},\bm{p})
\frac{1}{T}\int_{-T/2}^{T/2}\diff t\,
f_{\rm K}(\bm{x},\bm{p},t)
\eqe
If, in addition, we look at the radiation from a stationary
coarse-grained particle distribution $f(\bm{x},\bm{p})$, then,
replacing the time-averaged Klimontovich function by this distribution
leads to  
\eqb
\frac{\diff L}{\diff\omega}&=&
\int\diff^3\bm{x}\,\diff^3\bm{p}\,
P(\bm{x},\bm{p})
f(\bm{x},\bm{p})
\eqe
For the case of fluctuations in only the magnetic field, for example,
the particle energy is an integral of motion, and any homogeneous,
isotropic function of the Lorentz factor $\gamma(p)$ 
is a stationary solution of the
kinetic equation. Setting $f(\bm{x},\bm{p})=\frac{N}{V}\frac{1}{4\pi p^2}\delta(\gamma-\gamma(p))$,
we find
\eqb
\frac{\diff L}{\diff\omega}&=&
\frac{N}{4\pi V}
\int\diff^3\bm{x}\,\diff^2\bm{\Omega}\,
P(\bm{x},p\bm{\Omega})
\label{isoPower}
\eqe where $\bm{\Omega}=\bm{p}/p$. Thus, in order to compute the power
radiated per unit frequency interval, we must integrate the
instantaneous power over all directions of the velocity vector at each
point and over all positions within the source. In the following subsections
we present computations of the 
radiation produced from an ensemble of relativistic particles in 
static turbulent magnetic field configurations, employing a 
Monte Carlo integration of Equation (\ref{isoPower}).

\subsection{Emission spectrum}
\label{results_sect}

The character of the radiation produced by a relativistic particle depends on
whether the strength parameter 
\eqb
a=\frac{eF\lambda}{mc^2} 
\eqe 
is greater than or less than unity,
where $\lambda$ is the typical size of the field structures and
$eF=\langle\diff p_\bot/\diff t\rangle$ the average transverse force
on the particle \citep{landaulifshitz}. This Lorentz invariant
parameter is analogous to the strength parameter commonly used in
laser plasma physics, and is also sometimes called the 
\lq\lq wiggler\rq\rq\ or \lq\lq undulator\rq\rq\
parameter. For
static fields, it determines roughly the ratio of the deflection
angle to the beaming angle for a particle traversing a typical
structure. For simplicity, electric fields are neglected for the
remainder of this section ($F=B_\bot$).  Typically, the magnitude of
the strength parameter determines whether the
particle radiates in the synchrotron regime ($a>1$) or in the
so-called jitter/diffuse synchrotron regime ($a<1$).
 
For a given $B_\bot$ and $\lambda$, the maximum photon energy can be
determined.  However, the full details of the spectrum produced by a
particle, even in a relatively simple field configuration, can be
quite complicated.  Using the algorithm presented in the previous
section, the equations of motion can be integrated simultaneously with
equation (\ref{instpower2}), 
providing the complete spectrum.  For the results that
follow a fifth-order adaptive Runge-Kutta integrator was used
\citep{numericalrecipes}.  With the aid of some illustrative examples,
we demonstrate how different spectral features can be produced, and
emphasize the properties of the fields required to do so.

\subsection{Uniform fields -- the synchrotron approximation}
\begin{figure}
\vbox{
\includegraphics[width=0.45\textwidth]{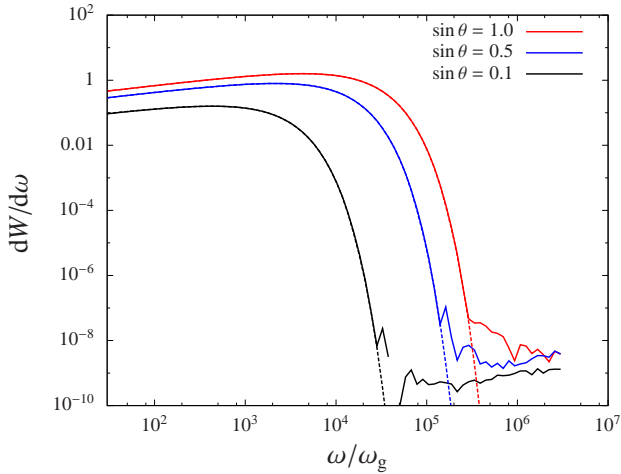}
}
\caption{ Instantaneous power spectrum produced by a particle of Lorentz factor
$\gamma=10^2$ with
  different pitch angles in a uniform magnetic field $\bm{B}$, as a function
of angular frequency in units of $\omega_{\rm g}=eB/mc$,  
with $\theta$ the
  angle between the particle velocity and the field. The numerical
  values (solid lines) are in excellent agreement with the instantaneous
  synchrotron approximation (dashed lines). The noise at high
  frequencies can be controlled by increasing both the number of terms 
  taken in the series and the integrator accuracy.
}
\label{fig1}

\end{figure}

As a first example, the radiation produced from a particle 
gyrating in a uniform field is compared to the analytic
solution for synchrotron radiation, Eq.~(\ref{instsyncheq}). 
The results are shown in
Fig. \ref{fig1} and are in excellent agreement with the analytic
result. At very high frequencies, the formation lengths become
extremely short, and $Q$ can be linear in $g$ for several periods.
Errors in the particle integrator can also become an issue.
For the results shown in Fig. \ref{fig1} a fractional
error control of $10^{-7}$ was used and the power was summed
from $n=0$ to $n=\pm10$. As discussed in section \ref{sect_algorithm},
the series fails to converge if $n^*>10$. However, this occurs only 
well above the cut-off frequency.
As we discuss in section \ref{PIC_sect}, when dealing with discrete 
time series, the \lq\lq synchrotron approximation\rq\rq\ must be taken at high
frequencies, where the formation length is small. 

The grouping
of the terms described in Equation (\ref{instpower}) is vital in
keeping the high frequency noise below the integrator accuracy. This 
is also important for calculations in turbulent fields when there is a large
variation in the roll-over frequency of the instantaneous power. Provided the
high frequency noise remains below the threshold, the results are
reliable.

\subsection{Turbulent fields}

In a turbulent magnetic field, the particle trajectory and resulting
radiation spectrum are generally quite complex
\citep{toptyginfleishman87}. Nevertheless, several qualitative
features can be understood in terms of the strength parameter,
although the product $B\lambda$ is replaced by a different value for each 
Fourier mode. There is now no single strength parameter
but rather a spectrum $a(k)=2\pi e B(k)/\left(mc^2 k\right)$, 
and radiation produced depends on several factors, most notably
the turbulent spectrum and the
magnitudes of $a(k_{\rm min})$ and $a(k_{\rm max})$.  

To investigate the effect of different turbulent field parameters,
static fields are constructed with the required properties. This is done
using a discrete Fourier transform description
following the method of \citet{giacalonejokipii99}. 
The magnetic field at a position $\bm{x}$ is 
$\bm{B}(\bm{x})=\bm{B_0}+\bm{\delta B}(\bm{x})$,
where $\bm{B}_0$ represents an external uniform mean field.
The turbulent field component 
is generated using $N$ Fourier modes, each with a random phase, direction and
polarization. In the
limit of large $N$,
\begin{equation}
\bm{\delta B}(\bm{x}) = \lim_{N\rightarrow\infty}
\sum_{n=1}^N A_n e^{{\rm i}(\bm{k}_n \cdot \bm{x} + \beta_n)} \bm{\hat{\xi}}_n
\end{equation}
represents an isotropic turbulent field.  Here $A_n$, $\beta_n$, 
$\bm{k}_n$ and $\bfm{\hat{\xi}}_n$ are the amplitude, phase, wave vector
and polarization vector for each mode $n$ respectively. The
polarization vector is determined by a single angle $0<\psi_n<2\pi$
\eqb 
\bfm{\hat{\xi}}_n = \cos \psi_n \bm{e}_x + {\rm i} \sin
\psi_n {\bf e}_y 
\eqe 
where $\bm{e}_x$ and $\bm{ e}_y$ are vectors,
orthonormal to $\bm{e}_z \equiv \bm{k}_n/k_n$. The vector $\bm{k}_n$ 
is determined by two additional angles, $0<\theta_n<\pi$ and
$0<\phi_n<2\pi$, and, for an isotropic distribution, should be
uniformly distributed on the unit sphere. These two angles
define a rotation matrix that determines $\bm{e}_x$ and
$\bm{e}_y$ \citep[e.g.][]{giacalonejokipii99}.  

The amplitude of each mode is 
\begin{equation}
 A_n^2 = \sigma^2 G_n 
\left[ \sum_{n=1}^N G_n\right]^{-1} \nonumber
\end{equation}
where the variance $\sigma^2$ is chosen such that the turbulent field is normalized
to give the required turbulence level:
\eqb
\eta&=&\frac{\left\langle\delta
    B^2\right\rangle}{B_0^2+\left\langle\delta B^2\right\rangle} \enspace.
\label{etadefinition}
\eqe

We use the following form for the power spectrum
\begin{equation}
G_n = \frac{\Delta V_n}{1+(k_nL_{\rm c})^\alpha}
\end{equation}
where $L_{\rm c}$ is the correlation length of the field 
and $\alpha$ is the asymptotic spectral index of the turbulence spectrum.
For the three-dimensional fields used in this paper
the normalization factor is $\Delta V_n = 4\pi k_n^2\Delta k_n$,
and the $\Delta k_n$ are chosen such that there is equal spacing in logarithmic $k$-space,
over the finite interval $k_{\rm min}\leq k\leq k_{\rm max}$.
For a detailed discussion of the statistical properties of fields constructed in this
manner see \citet{casseetal02}.
The field can be constructed at any point in space by summing over the $N$ modes,
providing an infinite spatial description of the fields.
This avoids the need for boundary conditions.
The parameters used for each field
construction are given in Table \ref{table1}.

%

\begin{deluxetable}{lcccccc}
\tablecaption{Turbulent Field Parameters}
\tablewidth{0.45\textwidth}
\tablehead{
Field & $B_{\rm rms}$ & $2\pi/k_{\rm max}$ & $2\pi/k_{\rm min}$ & $L_c$ & $\eta$ & $\alpha$}
\startdata
A          & $1.0$ & 2 & 160 & 80 & $1.0$ & 11/3\\
B          & $1.0$ & 2 & 320 & 160 & $1.0$ & 11/3\\
C          & $0.05$ & 0.5 & 10 & 5 & $1.0$ & 11/3\\
D          & $1.0$ & 0.05 & 10 & 5 & $1.0$ & 8/3\\
E          & $0.1$ & 0.1 & 1 & 0.5 & $1.0$ & 9/3\\
F          & $1.0$ & 0.05 & 10 & 5 & $0.1$ & 8/3\\
G          & $1.0$ & 0.1 & 10 & 5 & $0.9$ & 8/3\\
\enddata
\tablecomments{Parameters used in the field constructions for turbulent field spectra.
All quantities are dimensionless, with the magnetic field in units of  
an arbitrary normalization value $B_0$. All length scales are
in units of $mc^2/eB_0$. The maximum strength parameter in each run
is given approximately by the product $2\pi B_{\rm rms}/k_{\rm min}$.}
\label{table1}
\end{deluxetable}

\begin{figure}
\vbox{
\includegraphics[width=0.45\textwidth]{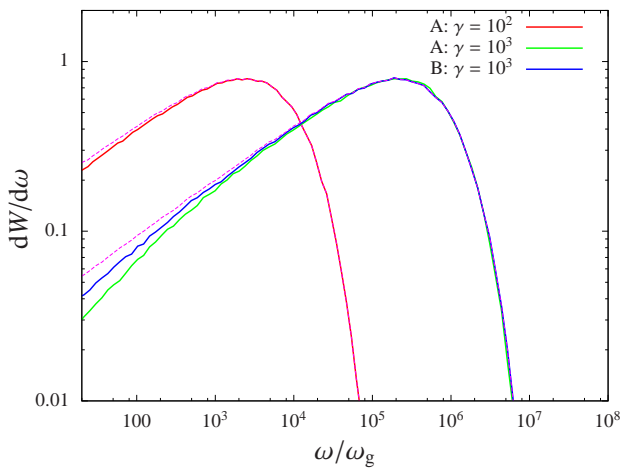}
}
\caption{ Radiation spectra for an isotropic distribution of 
mono-energetic particles
  in a fully turbulent field with strength parameter $a\gg1$,
as a function of frequency in units of $\omega_{\rm g}=eB_0/mc$. 
$\diff W/\diff \omega$ represents the average power emitted by 
each particle and is plotted in units $e^2\omega_{\rm g}/2\pi c$. 
The dashed lines show the integral of the instantaneous power, evaluated
in the synchrotron approximation.}
\label{fig2}

\end{figure}

The spectra are produced using a Monte Carlo integration of Equation
(\ref{isoPower}). At each frequency $\omega$ a sample 
particle of fixed Lorentz 
factor is placed
at a random location $\bm{x}_i$ inside a volume with dimensions several times the
size of the correlation length, $L_{\rm c}$, of the turbulent field.  To represent an
isotropic particle distribution, the particle is given a random
initial direction $\bm{\Omega}_i$, and the instantaneous power $P_i$ is
calculated. The average power emitted per particle at each frequency is determined using
a Monte Carlo integration:
\eqb
\frac{\diff W}{\diff\omega}
&=&
\frac{1}{4\pi V}\int_V\diff^3\bm{x}\,\int\diff^2\bm{\Omega}\,
P(\bm{x},p\bm{\Omega})
\nonumber\\
&\approx& \frac{1}{n}
\sum_{i=1}^n P_i
\label{isoPower2}
\eqe
The number of integration
points $n$ is determined from the condition that the standard
deviation error estimate is well below than $10\%$. This usually
requires only a relatively small number of points at low frequencies.
However, if $a(k_{\rm max})\ll1$, a
large number of points is needed at high frequencies 
 in order for the Monte Carlo
integrator to resolve these small scale structures.  For
comparison, the instantaneous synchrotron power, 
Equation~(\ref{instsyncheq}), is also calculated at each point.

Figures \ref{fig2} -- \ref{fig5} show the spectra produced by 
an isotropic homogeneous particle distribution in
turbulent isotropic fields with zero mean field component. A common
feature of each of these spectra is a hardening at low
frequencies. This arises because the particle begins to be deflected
by the turbulence through an angle comparable to that of the beaming cone
of the radiation, whilst traversing a photon formation length, which grows
towards low frequency. If the particle motion can be described as diffusion 
in the (small) angle $\theta$ between its velocity vector and a suitably chosen 
coordinate axis, 
this is known as the Landau-Pomeranchuk-Migdal (LPM) 
effect
\citep{landaupomeranchuk53b,landaupomeranchuk53a,migdal56} --- 
a well-studied phenomenon in the context of the suppression of 
bremsstrahlung and 
pair-production in crystals and other media \citep[for a review
see][]{klein99}, though not usually considered in the context of
synchrotron radiation 
\citep[although see][]{toptyginfleishman87,fleishman06}. 
The effect can be understood as follows:
For a
trajectory with constant curvature $\kappa$, 
and constant acceleration, $\ddot{\beta}=0$, the
particle displacement 
is $\Delta\approx\beta\tau
-c^2\kappa^2\tau^3/24$ (see equation \ref{displacement_appendix}). 
For low frequencies the resulting formation length 
is dominated by the $\tau^3$ term. It is this scaling that
gives synchrotron radiation its $\omega^{1/3}$ asymptote at low
frequencies. However, in turbulent fields, both the acceleration and 
curvature vary. Thus, at low frequencies, when the formation
lengths are long, a particle can undergo
multiple scattering within a formation time. 
In general, for small angle scattering ($a\ll\gamma$), the displacement is
$\Delta\approx\beta\tau-\frac{1}{2}\int_0^\tau \diff t \theta^2(t)+
\frac{1}{2\tau}\left(\int_0^\tau \diff t \theta(t)\right)^2$, and the spectrum should be
averaged over a large ensemble of particles 
\citep[e.g.][]{landaupomeranchuk53b, akhiezershulga87}.
For pitch-angle diffusion, i.e.\ $\left\langle\theta\right\rangle=0$ and
$\left\langle\theta^2\right\rangle \propto t$, the average displacement is proportional 
to $\tau^2$ at low frequencies, 
resulting in an $\omega^{1/2}$ spectrum. However, the transition to this regime requires
many scatterings and may not be realized within a formation length 
in a specific realization of a turbulent field. 

This is illustrated 
by the examples described in the following subsections.
The radiation spectra produced in these examples can be placed into
three broad categories, corresponding to the two extreme cases
where $a(k_{\rm min})\gg1$ or $a(k_{\rm min})\ll1$ and an intermediate
range in which $a(k_{\rm min})$ is of order unity.

\subsubsection{$a(k_{\rm min})\gg1$}

In the case of large strength parameters, since the particle in general 
sweeps through an angle larger than its beaming angle, the spectrum should 
resemble that of the instantaneous synchrotron spectrum close to the critical
frequency. Fig.~\ref{fig2}
shows the resulting spectrum for two different field configurations.
As expected, the spectrum matches very closely that of the instantaneous
synchrotron approximation in the vicinity of the roll-over frequency. 
Below this value, the numerically determined spectrum diverges slowly
from the instantaneous synchrotron line, becoming gradually harder at lower frequencies. 
Note that in the large strength parameter regime, the transition to the diffusive LPM
regime described above, i.e.\ the $\omega^{1/2}$ scaling,  
should occur when the formation length exceeds the longest wavelength
in the system, which occurs only at very low frequencies $\omega\sim\omega_{\rm c}/a(k_{\rm min})^3$.
The divergence from the synchrotron spectrum 
follows from Equation (\ref{displacement_appendix})
since now both $\dot{\kappa}$ and $\ddot{\beta}$ are non-zero, and the coefficient
of the $\tau^3$ term 
will have an additional time-dependence. 
For the range of frequencies considered, the spectrum does not 
approach a low-frequency
power-law asymptote, but continues to harden gradually as it approaches 
$\omega_{\rm g}$, where the synchrotron approximation fails and the beaming cone
is large.
The smaller the value of $a(k_{\rm min})$ the more rapidly
the spectrum diverges from that of the instantaneous 
synchrotron approximation. The spectra are not sensitive to the
value of $a(k_{\rm max})$ in the $a(k_{\rm min})\gg1$
regime, provided $a(k_{\rm max})\ll1$.

For frequencies above the roll-over frequency, as can be seen in 
Fig.~\ref{fig2}, the spectrum is in 
excellent agreement with the instantaneous synchrotron
approximation. On physical grounds it is expected that 
a power-law tail must occur at higher frequencies due to
the high frequency jittering resulting from modes with 
$a(k)<1$ \citep[see e.g.][]{fleishman06c}. 
However, for the turbulent spectra considered, the power 
associated with such fluctuations is extremely small, and
the numerical accuracy required to resolve such a feature
in the large $a(k_{\rm min})$ regime is beyond the capabilities of 
current computational resources.

\subsubsection{$a(k_{\rm min})\ll1$}

\begin{figure}
\vbox{
\includegraphics[width=0.45\textwidth]{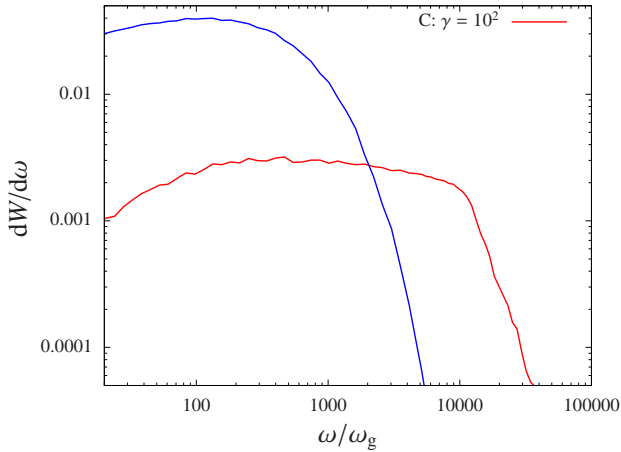}
}
\caption{ Radiation spectra for an isotropic distribution of particles
  in a fully turbulent field with all strength parameters $a<1$. The blue
  line is the integrated instantaneous synchrotron spectra. The high
  frequency asymptote is close to the shape of the turbulent spectrum
  $\propto\omega^{-11/3}$. The low frequency spectrum does not converge to
  a power law for the range of frequencies considered.}
\label{fig3}

\end{figure}

\begin{figure}
\vbox{
\includegraphics[width=0.45\textwidth]{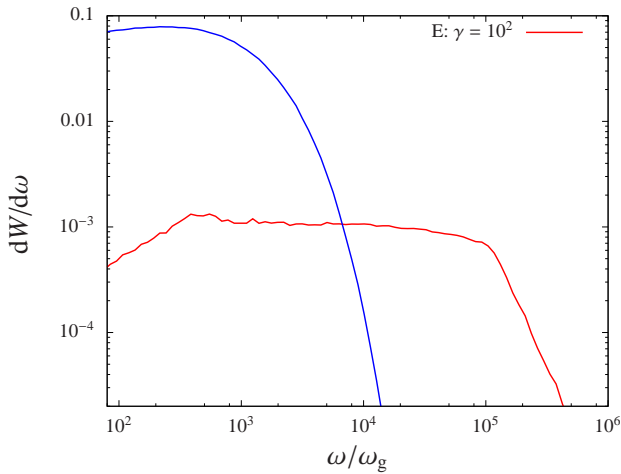}
}
\caption{ Radiation spectra for an isotropic distribution of particles
  in a fully turbulent field with strength parameters $a<1$ 
  using a smaller value for $a(k_{\rm min})$ and a
  larger dynamic range than in Fig. \ref{fig3}. The blue line is the
  integrated instantaneous synchrotron spectra. The high frequency
  asymptote is close to the shape of the turbulent spectrum
  $\propto\omega^{-3}$. The low frequency spectrum 
  has a spectral slope $\sim 0.7$. }
\label{fig5}

\end{figure}

For fields composed exclusively of small strength parameter fluctuations, 
the particle deflections are small and, at sufficiently high frequencies,
it is possible to use standard perturbation techniques
\citep{landaulifshitz,medvedev00,fleishman06}. Numerically, this regime
is far more challenging since the time steps in the integrator must resolve 
deflections in the particle's trajectory on the order $a(k_{\rm max})/\gamma$.
Figs.~\ref{fig3} and \ref{fig5} show the spectra produced in fields with
$a(k_{\rm min})=0.5$ and  $a(k_{\rm min})=0.1$ respectively.
Both spectra exhibit a
break at the critical frequency $\omega\approx\gamma^2k_{\rm min}c$.  
Above this frequency, the spectrum has a
power-law slope matching that of the turbulence spectrum. This can be
understood as the up-scattering of the virtual photons of the field
by the mono-energetic particles. In principle, the power law should extend up
to $\omega\approx\gamma^2 k_{\rm max}c$, however, the numerical accuracy 
of the 
integrator chosen for this example is insufficient to display the entire range.
At even higher frequencies, $\omega\gg\gamma^2k_{\rm max}c$  
the fields are constant
over the formation length of the particle, and the instantaneous
synchrotron approximation applies. As we discuss in section \ref{PIC_sect},
if the formation length of a particle is not well resolved, it is exactly in
this regime that the instantaneous synchrotron approximation must be used. 
 Below
the critical frequency, the photon formation time 
remains short compared to
the time taken to
deflect through an angle greater than $\gamma^{-1}$. The displacement
is approximately $\Delta\approx\beta\tau$ and, as in the case of relativistic 
bremsstrahlung, 
the spectrum is approximately flat $\diff W/\diff\omega\propto\omega^0$.  
Ultimately, at frequencies $\omega< a(k_{\rm min})\omega_{\rm c}\approx 
a(k_{\rm min})^2 \gamma^2 c k_{\rm min}$, the 
formation time exceeds the time needed to diffuse 
out of the beaming cone and the 
spectrum is determined by the LPM effect
\citep{fleishman06}.

\subsubsection{$a(k_{\rm min})\sim1$}
\begin{figure}
\centering
\vbox{
\includegraphics[width=0.45\textwidth]{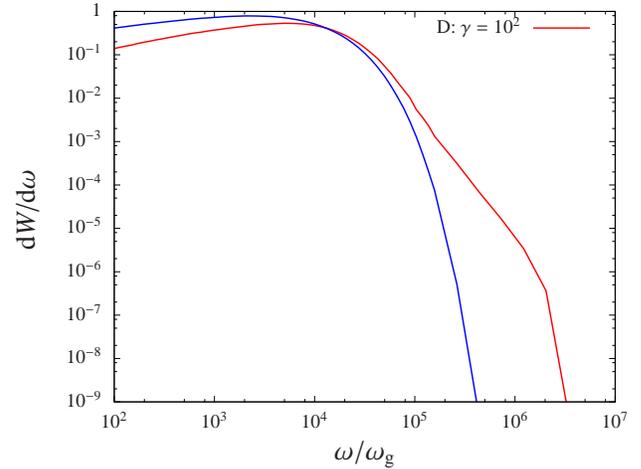}
}
\caption{ Radiation spectra for an isotropic distribution of particles
  in a fully turbulent field with strength parameter $a(k_{\rm min})\gtrsim1$
  and $a(k_{\rm max})\ll1$. The blue line is the
  integrated instantaneous synchrotron spectra. The index of the 
high frequency
  power-law component is close to that of the turbulent spectrum
  $\propto\omega^{-8/3}$. Evidence of a cut-off is observed close to where the 
  formation length $L_c\sim 1/k_{max}$, where the line must match up with the
  instantaneous synchrotron approximation. This cannot be resolved numerically.}
\label{fig4}

\end{figure}

The intermediate range where the value of $a(k_{\rm min})$ is somewhat larger than
unity, is interesting because it emerges from 
PIC simulations of Weibel mediated shocks
\citep[e.g][]{sironispitkovsky09b}.  An example of the spectrum
produced in such a field is shown in Fig. \ref{fig4}. 
For this example, the strength parameter $a(k_{\rm min})$
is of order unity, and the transition to the LPM regime occurs at
relatively high frequencies, close to the roll-over frequency.
However, above the roll-over frequency, unlike in the  $a(k_{\rm min})\gg1$ regime, 
the small strength parameter modes can be resolved, and similar to the
$a(k_{\rm min})\ll1$ spectra, a high frequency power-law emerges.
The shape of the power-law matches that of the turbulence spectrum. 
This presents a possible observational signature of
short wavelength turbulence at relativistic shocks. The presence
of such short wavelength turbulence is supported by current PIC simulations
in which Fermi acceleration is found to occur. 

\subsubsection{Non-zero mean field}

In general, the radiation spectrum can be affected 
if the mean field is non-zero
or if turbulence is generated on different scales such that the large
scale fluctuations act as a local mean field.
The latter situation could in principle be realized in the presence of large
scale MHD turbulence produced from interaction between the shock front
and density inhomogeneities \citep[e.g.][]{sironigoodman07} and short
wavelength turbulence produced via kinetic effects in the shock
transition region. In the presence of two populations of scatterers,
if they are generated on very different length-scales, it is possible
for the synchrotron
radiation of shock-accelerated particles to extend into the gamma-ray 
range, whereas for a single population of scatterers radiation losses restrict
it to relatively low frequency \citep{kirkreville10}.

\begin{figure}
\vbox{
\includegraphics[width=0.45\textwidth]{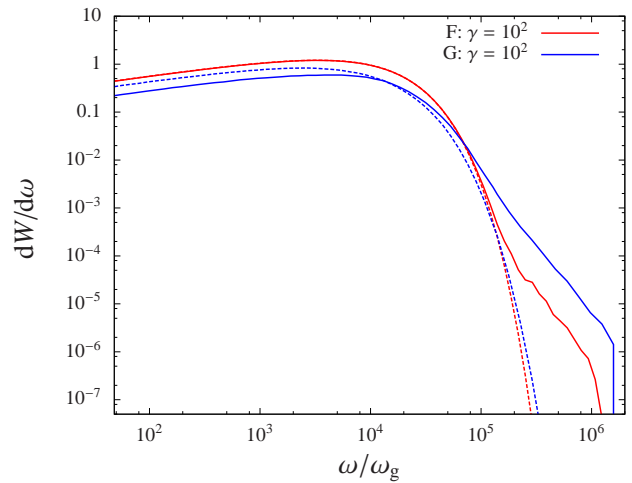}
}
\caption{ Spectra emitted in the presence of a finite mean field 
$\eta=0.1$ (red curves) and $\eta=0.9$ (blue curves), where $\eta$ is the 
ratio of the energy density in the turbulent field to the total energy 
density -- see (\ref{etadefinition}).  
The
  instantaneous synchrotron spectra are plotted using dashed
  lines. The total energy density in the magnetic field is fixed, so that
the magnitude of the average field differs in the two cases. 
As in Fig. \ref{fig4}, there is evidence of a cut-off at high frequencies.}
\label{fig6}

\end{figure}

Here we consider the radiation produced in a region with  
a mean field having a superimposed turbulence
spectrum.
If the energy in the turbulent fluctuations is
negligible with respect to the total field, 
$\eta\ll 1$, where $\eta$ is defined in (\ref{etadefinition}), 
the low frequency spectrum will match that of the instantaneous
synchrotron spectrum, since scattering will be ineffective, 
and to zeroth order, the particles simply gyrate about the mean field. 
At higher frequencies, provided modes with
$a<1$ exist, a power-law tail can emerge. Again, depending on the 
power associated with these modes, the numerical scheme can capture 
this feature, provided it is not too deep in the exponential cut-off region.
To illustrate this, we show in Figure \ref{fig6} the spectra produced in
turbulent fields, with modest maximum strength parameters, and different 
values of $\eta$. For small $\eta$, i.e.\ weak turbulence, the spectrum 
reproduces that of the instantaneous synchrotron spectrum, although 
a high frequency tail is also produced, due to the fluctuations on 
modes with $a<1$. As $\eta$ increases, more 
power goes into the high frequency emission, and a reduction in the
power at low frequencies is observed, although for the frequencies 
investigated, the spectrum maintains a $\omega^{1/3}$ scaling.

As the ratio of the energy density in the turbulent field to 
that of the total field is increased
further, we return to the previously investigated regimes. 
For example, in a Weibel mediated shock $1-\eta\ll1$ \citep{sironispitkovsky09a}. 
However, to investigate clearly identifiable 
signatures, it is necessary to move beyond the prescribed, 
homogeneous turbulent fields 
considered here to more self-consistent realizations,
resulting from the simulations.

\section{Trajectories and fields given as a time series}
\label{PIC_sect}

A PIC simulation is capable of producing a large 
number of time series listing the position and velocity
of the simulation particles and the values of the electromagnetic
fields at each time-step. Using these, it is possible to produce spectra
and light curves. 
The synthetic spectra presented in section \ref{turbulentfields} are
based on isotropic mono-energetic particle distributions as described
in section \ref{klim_sect}.  In general, however, the particle distribution is
not only energy dependent, but can be highly anisotropic.  This can in
principal be studied by numerically solving Equation (\ref{LWrad}) or
Equations (\ref{LWenergy}) and (\ref{pnot}) for each trajectory, and
then summing over trajectories, which is equivalent to integrating
over the particle distribution function.  However, the radiation from
an individual trajectory is beamed into an opening angle
$\sim1/\gamma$. If this is smaller than the scales on which the
particle distribution is anisotropic, the order of these operations
can be reversed \citep[see][section 3.2]{ginzburgsyrovatskii65}. The
average over the particle distribution is then replaced by an
integration over angles of the radiation emitted by a single
trajectory (which can be performed analytically), and the radiation
observed in a given virtual detector is given by summing over all
those trajectories whose velocity vector lies within the acceptance
cone of that detector. Formally,
\eqb 
\frac{\diff L}{\diff\omega\diff\bm{n}}&=&
 \int\diff^3\bm{x}\,\diff p\,p^2\,\diff^2\bm{\Omega}\, 
f(\bm{x},p\bm{\Omega},t)P(\bm{n},\omega,t) 
\nonumber\\
&\approx& 
 \int\diff^3\bm{x}\,\diff p\,p^2\, 
f(\bm{x},p\bm{n},t)
\int\diff^2\bm{\Omega}\,P(\bm{\Omega},\omega,t) \nonumber\\
&=& 
 \int\diff^3\bm{x}\,\diff p\,p^2\, 
f(\bm{x},p\bm{n},t)
P(\omega,t) 
\eqe
and the integrations over $\bm{x}$ and $p$ reduce in the PIC case to
summations over all trajectories that illuminate the specified detector.

For the high-energy emission of particles accelerated at a
relativistic shock front, the restriction imposed by this procedure is
not important, because the anisotropy of the particle distribution is
expected to be on a scale larger than the beaming angle. Thus, the
angular dependence of the emitted radiation found by
\citet{sironispitkovsky09b} and \citet{frederiksenetal10} 
should just reflect the
angular dependence of the distribution function at the relevant
particle energy, and would be preserved in this approach.

As pointed out by \citet{hededalphd}
the computation of synthetic spectra from trajectories 
taken from PIC simulations inevitably involves interpolation.
Specifically, the algorithm 
presented in (\ref{ptotal})--(\ref{p2}) transforms the
integration variable from time to phase. In order to split the 
contributions to the integral into
an
alternating series (\ref{instpower2}), 
the discrete trajectory must be interpolated. 

Interpolation is not a sensitive procedure provided 
many points are contained within a photon formation time,
a constraint that will be made more precise below.
An accurate evaluation of the instantaneous power at 
any time step can, for example, be obtained 
simply by linearly interpolating the functions $g$, $\gamma$, $\beta$,
$\dot{g}$, 
$\delta\bm{\beta}$ 
and $\delta\Delta$, which are known at all neighboring grid points.
When the photon formation length drops to 
only a few time steps, this procedure fails. However, the validity of 
the PIC simulation requires that the electromagnetic fields vary slowly between
time steps, which is precisely the condition for 
applicability of the generalized synchrotron formula (\ref{instsyncheq}). 
Therefore, in a valid simulation, 
the instantaneous power can safely be evaluated 
using this method, if the formation time is not long compared to the 
time step.
It follows that, for a given frequency, the method of evaluating the
instantaneous power at each of the discrete set of particle positions
$\bm{x}(t_n)$, depends on the value of the photon formation time at
that point.  

At high
frequencies, the formation time is short, and can be much shorter 
than the 
typical time-step used in PIC simulations, which is a fraction 
of a plasma cycle. 
It is straightforward to find for each time-step (labeled by $n$) the values
$\delta\Delta^\pm_n$ of the deviation of the displacement at the
neighboring points $n\pm1$. For a given frequency, the photon formation 
lengths in the forward and backward directions follow. Alternatively,
two critical frequencies $\omega^\pm_n$ can be found such that at
these frequencies the neighboring points lie precisely one formation
length away from $x_n$. From Equations (\ref{cohtime}) 
and (\ref{devdisplacement})
the critical frequencies are
\eqb
\omega^\pm_n=\frac{4\pi\gamma^2}{\delta t^\pm_n+2\gamma^2|
\delta\Delta^\pm_n|}
\eqe
where $\delta t^\pm_n=|t_{n\pm1}-t_n|$ is the time-step between neighboring
data points. 
If $\omega$ is close to or greater than $\omega^\pm_n$, 
then the coherence length is poorly resolved
and the synchrotron approximation must be used to compute the
instantaneous power. 
If, on the other hand, $\omega\ll\omega^\pm_n$,
then the coherence length is well resolved, and a numerical
integration is accurate.

The accuracy of the solution depends quite
strongly on the ability to resolve the peaks and troughs of the sine function
in Equation~(\ref{instpower2}).
PIC simulations usually work with a fixed time-step, in which 
case the resolution in
successive terms in (\ref{instpower2}) decreases. This case be seen by
considering the time evolution of the phase. Making a Taylor expansion
about the initial position gives 
\eqb
g\approx\omega\left[\frac{\tau}{2\gamma^2}+\frac{1}{24}c^2\beta^3\kappa^2\tau^3\right]
\eqe
For $\tau>3/\gamma c\kappa$, the $\tau^3$ term dominates and one can solve for $g=n\pi$ to give
\eqb
\tau_n\approx\frac{\gamma^2}{\omega_{\rm c}}\left(54 n \pi \frac{\omega_{\rm c} }{\omega}\right)^{1/3}
\eqe
It is readily seen that for larger $n$ the time interval between successive
integer multiples of the phase $g=n\pi$ decreases:
\eqb
\tau_{n+1}-\tau_{n} \approx \frac{\gamma^2}{\omega_{\rm c}}
\left(\frac{2 \pi}{n^2} \frac{\omega_{\rm c} }{\omega}\right)^{1/3}
\eqe
For this reason, it is essential to interpolate the functions
$Q$ and $g$, rather than the combination $Q \sin g$, 
and we have found that linear interpolation is adequate.
Then, trapezoidal integration in the phase 
$g$ is used with a maximum step-size of $\Delta g=2\pi/25$\footnote{
This is approximately the resolution required to 
calculate $\int_a^b\sin(x)\diff x$ to better than 
$99\%$ accuracy using trapezoidal
integration.} to evaluate the terms 
$n=-10\dots 10$ in (\ref{cohtime}).

\begin{figure}
\vbox{
\includegraphics[width=0.45\textwidth]{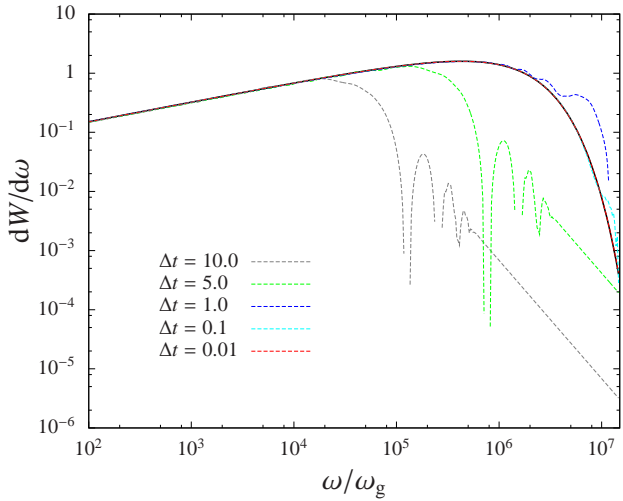}
}
\caption{
Synchrotron spectra found using discrete time-series 
data for a $\gamma=10^3$ particle
using linear interpolation on $Q$ and $g$ between
data points, for a range of time-step sizes (in units of $\omega_{\rm g}^{-1}$). 
In each case, the frequency at which the numerical result 
begins to deviate from the exact answer 
is in close agreement with (\ref{discom_max}).
}
\label{discretetest}

\end{figure}

As an illustrative example, we again consider the 
case of uniform circular motion. In 
Fig~\ref{discretetest} we plot the energy radiated per frequency interval
over one gyration. In addition to the analytic solution, 
the result of integrating the instantaneous
power calculated using various time-steps is shown. 
In this special example, both the instantaneous power 
and the frequencies $\omega^\pm_n$ are independent of time, 
so that use of the synchrotron approximation automatically yields the 
exact analytic result. 
The numerically 
determined power reproduces this result to within 1\% for frequencies 
\eqb
\label{discom_max}
\omega&<&\frac{1}{25}{\rm Min}\left({\omega^+_n,\omega^-_n}\right)\,.
\eqe
At higher frequencies, the instantaneous power itself
may still be evaluated accurately, since the interpolation scheme guarantees
25 points per photon formation length. However, since this quantity 
is evaluated only at each time-step,
the subsequent integration required to evaluate the 
radiated energy does not reach the required resolution. 

This suggests the following procedure when the algorithm is employed in 
an arbitrary field configuration:
The frequency at which the emission is to be evaluated, is 
compared at each time-step to the frequencies $\omega^\pm_n$. 
If $\omega$ satisfies the inequality (\ref{discom_max}), 
numerical integration is used.
Otherwise, the instantaneous synchrotron expression 
(\ref{instsyncheq}) is used.

An important property of this algorithm is that it 
avoids explicitly interpolating the particle's position and 
velocity. Such a procedure introduces discontinuities into the 
particle acceleration as a function of time, leading to 
high-frequency artifacts similar to those that arise when the 
acceleration of a hyperbolic
trajectory is abruptly terminated \citep{revillekirk10}.

As mentioned above,  the instantaneous power
can only be evaluated accurately by integrating over 
at least the 
first few formation lengths. The number of terms needed 
in (\ref{instpower2}) can be
considerably reduced with the aid of the Euler--van Wijngaarden
transform.  However, when calculating the total energy spectrum radiated by an
individual trajectory, it is also essential to resolve the
instantaneous power as a function of time. 
Given a finite time series of
positions and velocities, 
the radiation formulas apply
only if the trajectory is extrapolated ballistically outside of the finite
length time series, although the behavior in these regions does not affect
the results when frequencies
$\omega\gg\langle\gamma^2\rangle/T$ are considered. Here
$\langle\gamma^2\rangle$ is the average Lorentz factor squared along
the trajectory, and $T$ the total time. Of course, if a time-dependent
light-curve is to be generated, the restriction is much more severe, 
since then $T$ refers to the time-interval between successive evaluations.

To illustrate this, 
we consider the spectrum produced by a relativistic particle 
that undergoes an instantaneous scattering at $t=0$ through an angle $\alpha$.
The spectrum in this case is well known to be flat, $\omega^0$, at  
frequencies small compared to the inverse duration of the acceleration
\citep[for a detailed discussion see][chapter~37]{schwinger98}.
The angular integrated spectrum in this frequency range can be determined 
analytically \citep[e.g.][]{akhiezershulga87}
\eqb
\frac{\diff E}{\diff \omega}=
\frac{2 e^2}{\pi c}\left[
\frac{2\xi^2+1}{\xi\sqrt{\xi^2+1}}\ln|\xi+\sqrt{\xi^2+1}|-1\right]
\label{instscatspec}
\eqe

Although this example appears straightforward, it is, in fact, quite
demanding numerically, the reason for this being that the instantaneous power 
is itself an oscillatory function. The formation length at any given time $t$
can be easily calculated, but the exact expression is cumbersome.
Far from the scattering center, $t_{\rm c}= 4\pi\gamma^2/\omega$.
As the scatterer is approached, the formation length decreases, 
reaching a minimum at $t=0$, of
\eqb
l_{\rm c}=ct_{\rm c}\approx
\frac{4\pi\gamma^2c}{\omega(1+4\xi^2)}
\label{scat_coh}
\eqe
In this example there is no intrinsic time scale, so that 
we are free to choose arbitrary time, distance and frequency units. Defining 
a reference time unit $t_0$, we construct  dimensionless units $\hat{t}= t/t_0$,
$\hat{x}= x/ct_0$ and $\hat{\omega}= \omega t_0$.

\begin{figure}
\vbox{
\includegraphics[width=0.45\textwidth]{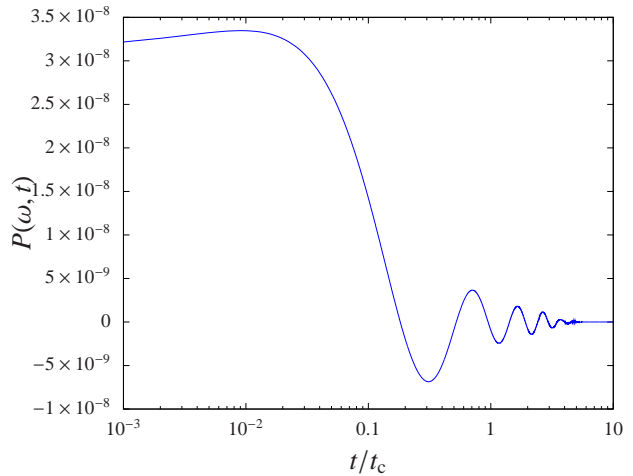}
}
\caption{
Instantaneous power at frequency $\omega=1$ (in arbitrary dimensionless
units) as a function of time produced by a particle 
with $\gamma=10^3$ that undergoes an instantaneous scattering at 
$t=0$ through an angle $\alpha=2\times10^{-4}$ ($\xi=0.1$).
Time is measured in units of the coherence time given by Eq.~(\ref{scat_coh}).
}
\label{scattest1}

\end{figure}

\begin{figure}
\vbox{
\includegraphics[width=0.45\textwidth]{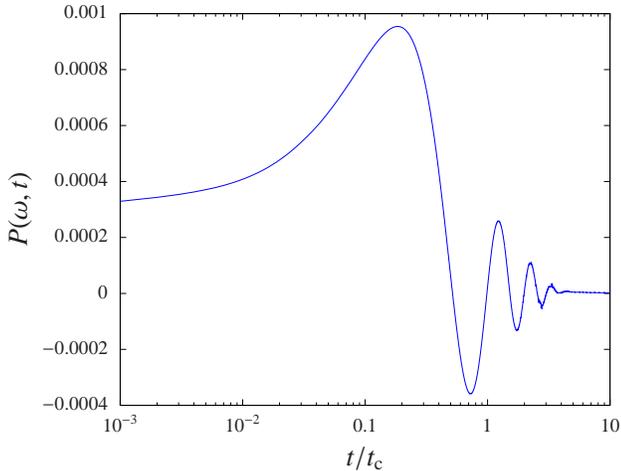}
}
\caption{
Instantaneous power at frequency $\omega=1$ 
as a function of time produced by a particle 
with $\gamma=10^3$ that undergoes an instantaneous scattering at 
$t=0$ through an angle $\alpha=2\times10^{-2}$ ($\xi=10$).
Time is measured in units of the coherence time given by Eq.~(\ref{scat_coh}).
}
\label{scattest2}

\end{figure}

Figures \ref{scattest1} and 
\ref{scattest2} show the instantaneous power as a 
function of time for $\xi=0.1$ and $\xi=10$, respectively.
The power oscillates with a slowly increasing period approximately equal to the 
formation time and damping with distance from the scatterer. 
In addition, there is an unresolved discontinuity 
at $t=0$, which arises because the particle velocity is also 
discontinuous at this point. However, after integration over $t$,
this feature has no influence on the energy radiated.
Clearly, the linear growth phase for $Q(g,t)$ will
increase with distance from the scattering event, and the number of terms in
the Euler--van Wijngaarden transform should be chosen such that the 
scattering is included. However, since the dominant contribution 
to the total energy radiated 
comes from the first few periods, the integral of the instantaneous
power converges rapidly and the error incurred from taking
only the first few formation lengths when calculating $P(\omega,t)$
is small.

\begin{figure}
\vbox{
\includegraphics[width=0.45\textwidth]{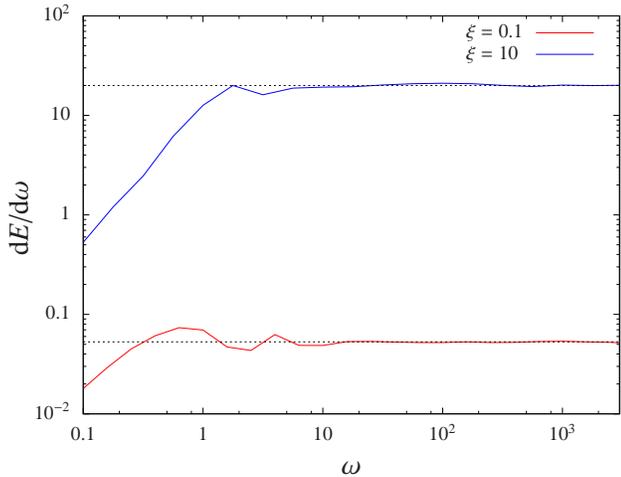}
}
\caption{
Plot demonstrating the low frequency limitation
due to finite endpoints of the trajectory. The dashed line is 
the analytic result from (\ref{instscatspec}). 
For the numerical evaluation, the
instantaneous power was evaluated on the time interval 
$-T<t<T$ where $T=\pi\gamma^2/(1+4\xi^2)$ with 
$\gamma=10^3$ and $\xi=0.1$ and $\xi=10$. 
}
\label{scat_spec}

\end{figure}

To demonstrate the effects of having a finite trajectory, we integrate
the instantaneous power over a time interval
\eqb
\frac{\diff E}{\diff\omega}=\int_{-T}^{T} P(\omega,t)\diff t\nonumber
\eqe
with $T={\pi\gamma^2}/{(1+4\xi^2)}$, corresponding to one
formation length for an emitted wave with frequency $\omega=4$. From figures
 \ref{scattest1} and \ref{scattest2}, it is clear that the solution will converge
only for frequencies much larger than this. The instantaneous power is integrated
using a finite time step trapezoidal integration 
for two different scattering angles $\xi=0.1$ and $\xi=10$, with $\gamma=10^3$,
as above.
The results are shown in figure \ref{scat_spec}. The result is in good agreement 
with the analytic solution above approximately $\omega=20$. This suggests that 
for a given trajectory, on a time interval $[-T,T]$, the minimum frequency that can 
be investigated must have at least 10 formation lengths in this time interval.

\section{Discussion}
\label{discussion}

In this paper we describe an algorithm for
calculating the radiation emitted by a relativistic charged particle 
moving in turbulent electromagnetic fields, and use it to investigate
the spectra that arise in a prescribed, stochastic realization of a 
static, turbulent magnetic field.
We also describe how to adapt the approach when the trajectory is
given as finite time series for the position, velocity and acceleration. 
The algorithm is based on formulating an \lq\lq instantaneous power\rq\rq\ 
at each point on the trajectory, and 
makes use of the concept of the photon formation length
to evaluate this quantity.  
It is suitable for use in post-processing 
the output from a particle-in-cell 
simulation.

Two problems arise with a numerical evaluation of the radiation. 
At high
frequencies, the finite time-resolution of the trajectory is a limitation. 
For relativistic particles, this problem can be alleviated by 
suitable grouping of the terms associated with the slowly and
rapidly varying components of the instantaneous power, which  
improves the stability and accuracy.
But even 
with the appropriate grouping of the terms, if the time-resolution
of the trajectory cannot be improved indefinitely, 
a purely numerical evaluation still fails at sufficiently high frequency. 
Fortunately, it is precisely in this range that the instantaneous power can safely 
be evaluated using the synchrotron approximation. Subsequent integration
of this quantity does not present a difficulty. 
However, for non-relativistic particles, or
for frequencies comparable to the instantaneous angular frequency of the
emitting particle, additional terms enter
into the expression for the 
instantaneous power (\ref{firstorderint})

At low frequencies, a
limitation is imposed when the trajectory is known only within a finite time
interval. This intrinsic 
restriction cannot be removed: the minimum frequency at which
a light curve can be computed is roughly $10\gamma^2/T$, where $T$ is the 
length of the available time series.   
In current PIC simulations, however, the neglect of 
the collective response of the plasma to the 
propagating waves, which leads to effects such as 
Razin-Tsytovich suppression and transition 
radiation, is also likely to be important.
These should  
intervene at frequencies below roughly 
$\sim\gamma\omega_{\rm p}$, where $\omega_{\rm p}=\sqrt{4\pi n e^2/m}$ is
the plasma frequency and $n$ the number density, 
but to date it does not appear feasible 
to account for such effects self-consistently.
For particles of Lorentz factor $100$, these estimates imply
that the intrinsic restriction is more important than the neglect of collective 
effects only when time-series shorter than about $10^3$ plasma cycles are used 
to compute the emitted radiation.

Our results obtained for 3D magnetostatic
turbulence confirm that one observational signature of short
length-scale turbulence, is the presence of a high-frequency power-law
tail in the mono-energetic emission spectrum
\cite{fleishman06b}.  For a power law of
electrons $\diff n/\diff\gamma \propto \gamma^{-p}$, one expects a
synchrotron 
power-law spectrum $F_{\omega}\propto \omega^{-s}$, where
$s=(p-1)/2$, for frequencies below the roll-over frequency of
the maximum energy electrons.  
Observations of GRBs place this index
in the range $2<p<2.8$, corresponding to a spectral index of
$0.5<s<0.9$. Thus, unless the turbulence index is extremely hard,
$\alpha<1$, the photon spectrum will not harden at high frequencies, and
this observational signature may be difficult to distinguish from a cut-off.

The synchrotron spectrum of particles radiating in a uniform field is
nowhere harder than an $\omega^{1/3}$ power law, which, 
since harder spectra have been observed in gamma-ray bursts, 
has led to the discussion of a synchrotron \lq line
of death\rq\ \citep{preeceetal98}.  
Our results confirm that this is generally true for isotropic particle
distributions in large scale, static, 3D turbulence. 
However, in agreement with other treatments
\citep{fleishmanurtiev10}
we find that, in the presence of
large amplitude turbulence on short length-scales, the low-frequency
asymptote can diverge from this value. 
We find low frequency spectra that are typically harder than $\omega^{1/3}$.
For fields with $a\gg1$ the spectrum exhibits a gradual hardening with the
slope increasing approximately $0.05$ per decade in frequency. For fields with
strength parameters $a\gtrsim1$, low frequency power-law asymptotes are produced, 
$F_{\omega}\propto\omega^q$, with $1/3\leq q\leq 1/2$. For fields with $a<1$
slightly larger values of $q$ appear to be
possible, although shocks with such small strength parameters are poor
accelerators \citep{kirkreville10}. Spectra as hard as $\omega^1$ 
are known to be produced by 
weak ($a\ll1$) turbulence that can be factorized into 
2D and 1D components \citep{fleishman06,medvedev06}. 
However, they do not arise in 
our results, which are based on a fully 3D~turbulence model.

\acknowledgements{
We thank A.~M. Taylor and S. O'Sullivan for helpful discussions. 
B.R. gratefully acknowledges support from the Alexander von Humboldt foundation. }


\appendix

\section{Synchrotron emission}
\label{appendix_synchrotron}
We start by making a Taylor expansion of the particle position, the 
deviations from ballistic motion and the phase. This can be
achieved using a purely geometric description of the trajectory.

First, define the tangent, normal and binormal
unit-vectors:
\eqb
\bm{T}(t)&=&\frac{\bm{\beta}(t)}{\beta(t)} 
\nonumber\\
\bm{N}(t)&=&\frac{\diff\bm{T}/\diff t}
{\left|\diff\bm{T}/\diff t\right|}\,=\,\frac{\diff\bm{T}/\diff t}{c\beta(t)\curvature(t)}
\nonumber\\
\bm{B}(t)&=&\bm{T}(t)\wedge\bm{N}(t)
\eqe
The quantity $\curvature(t)$ is called the curvature of the trajectory.
The Frenet-Serret formulae give the evolution of these vectors 
along the trajectory:
\eqb
\left(
\begin{array}{c}
\diff \bm{T}/\diff t\\
\\
\diff \bm{N}/\diff t\\
\\
\diff \bm{B}/\diff t
\end{array}\right)
&=&
\left(
\begin{array}{ccc}
   0  &   c\beta\curvature   &   0\\ 
&&\\
 -c\beta\curvature  &   0   &   c\beta\torsion\\ 
&&\\
  0  &   -c\beta\torsion   &   0 
\end{array}
\right)
\left(
\begin{array}{c}
\bm{T}\\
\\
\bm{N}\\
\\
\bm{B}
\end{array}\right)
\eqe
where $\torsion$ is called the torsion of the trajectory.

Therefore,
\eqb
\delta\bm{R}(t,\tau)&=&
\left[\frac{\tau^2}{2}
c\dot{\beta}+\frac{\tau^3}{6}\left(c\ddot{\beta}-
c^3\beta^3\curvature^2\right)\right]\bm{T}
\nonumber\\
&&+
\left[\frac{\tau^2}{2}c^2\beta^2\curvature+\frac{\tau^3}{6}\left(
3c^2\beta\dot{\beta}\curvature+c^2\beta^2\dot{\curvature}\right)\right]\bm{N}
+
\frac{\tau^3}{6}c^3\beta^3\curvature\torsion\bm{B}+\textrm{O}(\tau^4)
\\
\delta\bm{\beta}(t,\tau)&=&
\left[\tau\dot{\beta}+\frac{\tau^2}{2}\left(\ddot{\beta}
-c^2\beta^3\curvature^2\right)\right]\bm{T}
+
\left[\tau c\beta^2\curvature+\frac{\tau^2}{2}\left(3c\beta\dot{\beta}\curvature+
c\beta^2\dot{\curvature}\right)\right]\bm{N}
\nonumber\\
&&+\frac{\tau^2}{2}c^2\beta^3\curvature\torsion\bm{B}+\textrm{O}(\tau^3)
\eqe
Then, using 
\eqb
\bm{\beta}\cdot\delta\bm{R}/c&=&
\frac{\tau^2}{2} \beta\dot{\beta}+\frac{\tau^3}{6}
\left(\beta\ddot{\beta}-c^2\beta^4\curvature^2\right)
+\textrm{O}(\tau^4)
\\
\left(\delta\bm{R}\right)^2/c^2&=&
\frac{\tau^4}{4}\left(\dot{\beta}^2+c^2\beta^4\curvature^2\right)
+\textrm{O}(\tau^5)
\eqe
one finds

\eqb
\delta\Delta(t,\tau)&=&
\left[\tau^2\beta^2+2\tau\bm{\beta}\cdot\delta\bm{R}/c+
\left(\delta\bm{R}\right)^2\right/c^2]^{1/2}-|\tau|\beta
\\
&=&\frac{\tau|\tau|}{2}\dot{\beta}
+
\frac{\left|\tau^3\right|}{24}
\left( 4\ddot{\beta}-c^2\beta^3\curvature^2\right)
+\textrm{O}(\tau^4)
\label{displacement_appendix}
\eqe
Substituting into the definition of the phase-lag:
\eqb
g(t,\tau)&=&
\omega\tau\left[1-\beta-\frac{\tau}{2}\dot{\beta}
-\frac{\tau^2}{24}\left(4\ddot{\beta}-c^2\beta^3\curvature^2\right)
\right]+\textrm{O}(\tau^4)
\\
\dot{g}&=&
\omega\left[1-\beta-\tau\dot{\beta}
-\frac{\tau^2}{8}\left(4\ddot{\beta}-c^2\beta^3\curvature^2\right)
\right]+\textrm{O}(\tau^3)
\eqe

At this point, two additional assumptions are introduced:
\begin{enumerate}
\item
the electromagnetic fields are constant over a photon formation length,
i.e., $\ddot{\beta}=0$
\item
{\em linear acceleration emission} is negligible, i.e., $\dot{\beta}=0$
\end{enumerate}
The first is an implicit condition for the validity of a PIC simulation, when 
the photon formation time is comparable or shorter than the time step. The 
second is fulfilled under normal 
conditions ($\left|E\right|\lesssim\left|B\right|$).

Then, writing
\eqb
x&=&\frac{c\curvature\gamma\tau}{2}
\\
\omega_{\rm c}&=&\frac{3}{2}\gamma^3 c\curvature
\eqe
one finds
\eqb
g(t,\tau)&\approx&\omega\tau\left[1-\beta+\frac{1}{24}c^2\beta^3\curvature^2\tau^2\right]
\nonumber\\
&\approx&\frac{3\omega}{2\omega_{\rm c}}\left[x+\frac{x^3}{3}\right]
\\
\eqe
and 
\eqb
\dot{g}(t,\tau)&\approx&\frac{3\gamma\omega c\curvature}{4\omega_{\rm c}}
\left(1+x^2\right)
\eqe
At the frequencies of interest ($\omega\sim\omega_{\rm c}$), 
the dominant contribution to the integrals, which
arises for $g\sim1$, occurs for $x\sim1$. In this case, the 
the contribution of the first two non-vanishing terms 
in the Taylor expansions of 
both $g$ and $\dot{g}$ are comparable,  Thus, in expanding the integrands in 
(\ref{instpower}) it is necessary to include both these terms, whereas
the lowest order non-vanishing contributions to $\delta\Delta$
and $\bm{\beta}\cdot\delta{\bm{\beta}}$ are sufficient. 

The instantaneous power (\ref{instpower}) 
is then
\eqb
P(\omega,t)&\approx&
\frac{e^2\omega}{\pi c}
\frac{4}{3\gamma^2}
\int_{0}^\infty \diff x\, 
\frac{\left(x^2+\frac{x^4}{2}\right)}%
{x+\frac{x^3}{3}}\sin\left[\frac{3\omega}{2\omega_{\rm c}}\left(
x+\frac{x^3}{3}\right)\right]
\eqe
Writing $\eta=\omega/\omega_{\rm c}$,
\eqb
\frac{\diff}{\diff\eta}\left(\frac{1}{\eta}P\right)&=&
\frac{e^2}{\pi c}
\frac{2\omega_{\rm c}}{\gamma^2}
\int_0^\infty\diff x\,\left(x^2+\frac{x^4}{2}\right)\cos\left[\frac{3\eta}{2}
\left(x+\frac{x^3}{3}\right)\right]
\eqe
Then, using 
\eqb
\frac{1}{\sqrt{3}}K_{2/3}(\eta)&=&\int_0^\infty\diff x\,x
\sin\left[\frac{3\eta}{2}\left(x+\frac{x^3}{3}\right)\right]
\eqe
to find
\eqb
{\cal I}_2\,\equiv\,\int_0^\infty\diff x\,x^2
\cos\left[\frac{3\eta}{2}\left(x+\frac{x^3}{3}\right)\right]
&=&\frac{1}{\sqrt{3}}\left[K_{2/3}'(\eta)+\frac{2}{3\eta}K_{2/3}(\eta)\right]
\\
{\cal I}_4\,\equiv\,
\int_0^\infty\diff x\,x^4
\cos\left[\frac{3\eta}{2}\left(x+\frac{x^3}{3}\right)\right]
&=&\frac{2}{\sqrt{3}}K_{2/3}'(\eta)-3{\cal I}_2
\eqe
and noting that 
\eqb
\frac{\diff}{\diff\eta}K_{2/3}(\eta)-\frac{2}{3\eta}K_{2/3}(\eta)&=&
-K_{5/3}(\eta)
\eqe
one arrives at the standard expression for angle-integrated synchrotron 
radiation
\eqb
P(\eta,t)&=&
\frac{\sqrt{3}e^2\gamma\curvature}{2\pi}
\eta\int_\eta^\infty\diff x\,K_{5/3}(x)
\eqe


\begin{thebibliography}{33}
\expandafter\ifx\csname natexlab\endcsname\relax\def\natexlab#1{#1}\fi

\bibitem[{{Achterberg} {et~al.}(2001){Achterberg}, {Gallant}, {Kirk}, \&
  {Guthmann}}]{achterbergetal01}
{Achterberg}, A., {Gallant}, Y.~A., {Kirk}, J.~G., \& {Guthmann}, A.~W. 2001,
  \mnras, 328, 393

\bibitem[{{Akhiezer} \& {Shul'ga}(1987)}]{akhiezershulga87}
{Akhiezer}, A.~I., \& {Shul'ga}, N.~F. 1987, Soviet Physics Uspekhi, 30, 197

\bibitem[{{Casse} {et~al.}(2002){Casse}, {Lemoine}, \&
  {Pelletier}}]{casseetal02}
{Casse}, F., {Lemoine}, M., \& {Pelletier}, G. 2002, \prd, 65, 023002

\bibitem[{{Derishev}(2007)}]{derishev07}
{Derishev}, E.~V. 2007, \apss, 309, 157

\bibitem[{{Fleishman}(2006{\natexlab{a}})}]{fleishman06b}
{Fleishman}, G.~D. 2006{\natexlab{a}}, \mnras, 365, L11

\bibitem[{{Fleishman}(2006{\natexlab{b}})}]{fleishman06}
---. 2006{\natexlab{b}}, \apj, 638, 348

\bibitem[{{Fleishman}(2006{\natexlab{c}})}]{fleishman06c}
{Fleishman}, G.~D. 2006{\natexlab{c}}, in Lecture Notes in Physics, Berlin
  Springer Verlag, Vol. 687, Geospace Electromagnetic Waves and Radiation, ed.
  {J.~W.~Labelle \& R.~A.~Treumann}, 87--+

\bibitem[{{Fleishman} \& {Urtiev}(2010)}]{fleishmanurtiev10}
{Fleishman}, G.~D., \& {Urtiev}, F.~A. 2010, \mnras, 406, 644

\bibitem[{{Frederiksen} {et~al.}(2010) {Trier Frederiksen},
  {Haugboelle}, {Medvedev}, \& {Nordlund}}]{frederiksenetal10}
{Trier Frederiksen}, J., {Haugb{\o}lle}, T., {Medvedev}, M.~V.,  \& {Nordlund}, {\AA}.
  2010, ArXiv e-prints

\bibitem[{{Giacalone} \& {Jokipii}(1999)}]{giacalonejokipii99}
{Giacalone}, J., \& {Jokipii}, J.~R. 1999, \apj, 520, 204

\bibitem[{{Ginzburg} \& {Syrovatskii}(1965)}]{ginzburgsyrovatskii65}
{Ginzburg}, V.~L. \& {Syrovatskii} S.~I., \araa, 3, 297

\bibitem[{{Hededal}(2005)}]{hededalphd}
{Hededal}, C. 2005, PhD thesis, , Niels Bohr Institute

\bibitem[{{Kirk} \& {Reville}(2010)}]{kirkreville10}
{Kirk}, J.~G., \& {Reville}, B. 2010, \apjl, 710, L16

\bibitem[{{Klein}(1999)}]{klein99}
{Klein}, S. 1999, Reviews of Modern Physics, 71, 1501

\bibitem[{{Landau} \& {Lifshitz}(1971)}]{landaulifshitz}
{Landau}, L.~D., \& {Lifshitz}, E.~M. 1971, {The classical theory of fields},
  ed. {Landau, L.~D.~\& Lifshitz, E.~M.}

\bibitem[{{Landau} \& {Pomeranchuk}(1953{\natexlab{a}})}]{landaupomeranchuk53b}
{Landau}, L.~D., \& {Pomeranchuk}, I. 1953{\natexlab{a}}, Dokl. Akad. Nauk Ser.
  Fiz., 92, 735

\bibitem[{{Landau} \& {Pomeranchuk}(1953{\natexlab{b}})}]{landaupomeranchuk53a}
---. 1953{\natexlab{b}}, Dokl. Akad. Nauk Ser. Fiz., 92, 535

\bibitem[{{Martins} {et~al.}(2009{\natexlab{a}}){Martins}, {Martins},
  {Fonseca}, \& {Silva}}]{jmartinsetal09}
{Martins}, J.~L., {Martins}, S.~F., {Fonseca}, R.~A., \& {Silva}, L.~O.
  2009{\natexlab{a}}, in Society of Photo-Optical Instrumentation Engineers
  (SPIE) Conference Series, Vol. 7359, Society of Photo-Optical Instrumentation
  Engineers (SPIE) Conference Series

\bibitem[{{Martins} {et~al.}(2009{\natexlab{b}}){Martins}, {Fonseca}, {Silva},
  \& {Mori}}]{martinsetal09}
{Martins}, S.~F., {Fonseca}, R.~A., {Silva}, L.~O., \& {Mori}, W.~B.
  2009{\natexlab{b}}, \apjl, 695, L189

\bibitem[{{Medvedev}(2000)}]{medvedev00}
{Medvedev}, M.~V. 2000, \apj, 540, 704

\bibitem[{{Medvedev}(2006)}]{medvedev06}
---. 2006, \apj, 637, 869

\bibitem[{{Medvedev} {et~al.}(2010){Medvedev}, {Trier Frederiksen},
  {Haugboelle}, \& {Nordlund}}]{medvedevetal10}
{Medvedev}, M.~V., {Trier Frederiksen}, J., {Haugb{\o}lle}, T., \& {Nordlund}, {\AA}.
  2010, ArXiv e-prints

\bibitem[{{Melrose}(1978)}]{melrose78}
{Melrose}, D.~B. 1978, \apj, 225, 557

\bibitem[{{Migdal}(1956)}]{migdal56}
{Migdal}, A.~B. 1956, Physical Review, 103, 1811

\bibitem[{{Preece} {et~al.}(1998){Preece}, {Briggs}, {Mallozzi}, {Pendleton},
  {Paciesas}, \& {Band}}]{preeceetal98}
{Preece}, R.~D., {Briggs}, M.~S., {Mallozzi}, R.~S., {Pendleton}, G.~N.,
  {Paciesas}, W.~S., \& {Band}, D.~L. 1998, \apjl, 506, L23

\bibitem[{{Press} {et~al.}(1986){Press}, {Flannery}, \&
  {Teukolsky}}]{numericalrecipes}
{Press}, W.~H., {Flannery}, B.~P., \& {Teukolsky}, S.~A. 1986, {Numerical
  recipes. The art of scientific computing}, ed. {Press, W.~H., Flannery,
  B.~P., \& Teukolsky, S.~A.}

\bibitem[{{Reville} \& {Kirk}(2010)}]{revillekirk10}
{Reville}, B., \& {Kirk}, J.~G. 2010, \apj, 715, 186

\bibitem[{{Schwinger}(1949)}]{schwinger49}
{Schwinger}, J. 1949, Physical Review, 75, 1912

\bibitem[{{Schwinger} {et~al.}(1998){Schwinger}, {DeRead}, {Milton}, \&
  y~{Tsai}}]{schwinger98}
{Schwinger}, J., {DeRead}, L.~L., {Milton}, K.~A., \& y~{Tsai}, W. 1998,
  Classical Electrodynamics (Perseus Books, Reading)

\bibitem[{{Sironi} \& {Goodman}(2007)}]{sironigoodman07}
{Sironi}, L., \& {Goodman}, J. 2007, \apj, 671, 1858

\bibitem[{{Sironi} \& {Spitkovsky}(2009{\natexlab{a}})}]{sironispitkovsky09a}
{Sironi}, L., \& {Spitkovsky}, A. 2009{\natexlab{a}}, \apj, 698, 1523

\bibitem[{{Sironi} \& {Spitkovsky}(2009{\natexlab{b}})}]{sironispitkovsky09b}
---. 2009{\natexlab{b}}, \apjl, 707, L92

\bibitem[{{Spitkovsky}(2005)}]{spitkovsky05}
{Spitkovsky}, A. 2005, in American Institute of Physics Conference Series, Vol.
  801, Astrophysical Sources of High Energy Particles and Radiation, ed.
  {T.~Bulik, B.~Rudak, \& G.~Madejski}, 345--350

\bibitem[{{Spitkovsky}(2008{\natexlab{a}})}]{spitkovsky08a}
{Spitkovsky}, A. 2008{\natexlab{a}}, \apjl, 673, L39

\bibitem[{{Spitkovsky}(2008{\natexlab{b}})}]{spitkovsky08b}
---. 2008{\natexlab{b}}, \apjl, 682, L5

\bibitem[{{Toptygin} \& {Fleishman}(1987)}]{toptyginfleishman87}
{Toptygin}, I.~N., \& {Fleishman}, G.~D. 1987, \apss, 132, 213

\end{thebibliography}
\end{document}